\documentclass[prd,floatfix,showpacs]{revtex4}
\usepackage{epsfig}

\newcommand{\eq}[1]{Eq.~(\ref{#1})}

\newcommand{\be}{\begin{equation}}
\newcommand{\ee}{\end{equation}}
\newcommand{\bea}{\begin{eqnarray}}
\newcommand{\eea}{\end{eqnarray}}
\newcommand{\ben}{\begin{eqnarray*}}
\newcommand{\een}{\end{eqnarray*}}

\newcommand{\e}{\varepsilon}

\newcommand{\ga}{\gamma}

\newcommand{\de}{\delta}

\newcommand{\si}{\sigma}

\newcommand{\ro}{\rho}
\newcommand{\la}{\lambda}

\newcommand{\ta}{\tau}

\newcommand{\pd}{\partial}
\newcommand{\vph}{\varphi}
\renewcommand{\th}{\theta}
\newcommand{\cd}{{\cal D}}
\newcommand{\cs}{{\cal S}}
\renewcommand{\div}{\vec{\nabla}}

\newcommand{\s}[2]{{#1}\!\cdot\!{#2}}
\newcommand{\ov}[1]{\overline{#1}}

\newcommand{\dx}[1]{d^4{#1}\,}

\begin{document}
\title{Resolving temporal Gribov copies in Coulomb gauge Yang--Mills theory}
\author{H.~Reinhardt}
\author{P.~Watson}
\affiliation{Institut f\"ur Theoretische Physik, Universit\"at T\"ubingen, 
Auf der Morgenstelle 14, D-72076 T\"ubingen, Deutschland}
\begin{abstract}
The continuum Yang-Mills functional integral within the first order formalism 
and in Coulomb gauge is studied.  In particular, the temporal zero-modes of 
the Faddeev-Popov operator are explicitly accounted for.  It is shown that 
the treatment of these zero-modes results in the constraint that the total 
color charge of the system vanishes at all times.  Further, it is argued 
that the functional integral is effectively fully gauge-fixed once Gauss' 
law has been resolved in Coulomb gauge.
\end{abstract}
\maketitle
\section{Introduction}
\setcounter{equation}{0}

Coulomb gauge quantum chromodynamics [QCD] and Yang--Mills theory have 
recently begun to attract considerable attention, primarily because of their 
great potential in studying confinement.  This potential has long been 
recognized and led to the Gribov--Zwanziger scenario for confinement 
\cite{Gribov:1977wm,Zwanziger:1995cv,Zwanziger:1998ez} whereby the temporal 
component of the gluon propagator provides for a long-range confining force 
whilst the transverse spatial components are suppressed in the infrared.  
However, progress in Coulomb gauge has been hindered by the inherent 
noncovariance of the gauge condition.  Gratifyingly, the technical obstacles 
are being steadily overcome.

Various approaches to Coulomb gauge are currently being considered.  There 
exist lattice studies \cite{Burgio:2008jr,Quandt:2007qd,Voigt:2007wd,
Cucchieri:2006hi,Langfeld:2004qs,Cucchieri:2000gu} that are beginning to 
shed light on the nonperturbative behavior of the propagators.  Also, a 
Hamiltonian-based approach to the problem \cite{Szczepaniak:2001rg,
Szczepaniak:2003ve,Feuchter:2004mk,Reinhardt:2004mm} based on the original 
work of Ref.~\cite{Christ:1980ku} describes various features of the system 
in a coherent way \cite{Reinhardt:2007wh,Reinhardt:2008ek,Campagnari:2008yg}.  
The Lagrange-based (Dyson--Schwinger) approach to the problem 
\cite{Watson:2006yq,Zwanziger:1998ez} is also making progress.  One common 
theme to all these studies (and central to the Gribov--Zwanziger confinement 
scenario) is the importance of the Faddeev--Popov operator that arises after 
one fixes the gauge.

In addition to its noncovariant nature, Coulomb gauge is also incomplete --- 
the gauge is only partially fixed.  After applying Coulomb gauge, one can 
still perform time-dependent (spatially independent) gauge transformations and 
this raises potential questions about the strict validity of Coulomb gauge 
\cite{Baulieu:1998kx}.  On the other hand, trying to completely fix the 
gauge in the continuum seems to lead to a contradiction \cite{Watson:2007fm}.  
Related to the gauge-fixing and noncovariance issues in Coulomb gauge is the 
problem of energy divergences.  Because the gauge-fixing only involves 
spatial derivatives, the ghost propagator is independent of the energy and 
closed loops involving these ghosts appear at first sight to exhibit a pure 
(and unregularizable) energy divergence.  It has been argued that these 
divergences cancel to all orders perturbatively \cite{Niegawa:2006hg}.  
This cancellation, which has been explicitly verified at one-loop order 
\cite{Watson:2007mz,Watson:2007vc}, is not arbitrary: within the first order 
formalism, the Faddeev--Popov determinant cancels when the temporal gauge 
field is integrated out \cite{Watson:2006yq,Zwanziger:1998ez}.

In this paper, the Faddeev--Popov operator is considered with the focus on 
the temporal zero-modes.  These modes embody both the noncovariant and 
incomplete aspects of the Coulomb gauge-fixing.  The first order, 
(Lagrange-based) functional formalism in the continuum is employed.  It is 
shown that after Gauss' law is implemented (which in the functional approach 
arises after integrating out the temporal gauge field), the temporal 
zero-modes can be accounted for and their resolution leads to a completely 
gauge-fixed action that is free of energy divergences (in the sense that the 
Faddeev--Popov determinant cancels) with a well-defined, conserved and 
vanishing total color charge.

\section{Functional formalism}
\setcounter{equation}{0}
The starting point for this study is the functional integral associated with 
continuum Yang-Mills theory (see Ref.~\cite{Watson:2006yq} for a complete 
description),
\be
Z=\int\cd\Phi\exp{\left\{\imath\cs_{YM}\right\}},\;\;\cs_{YM}
=\frac{1}{2}\int\dx{x}\left[E^2-B^2\right],
\ee
with the chromoelectric and chromomagnetic fields given in terms of the 
spatial ($\vec{A}$) and temporal ($\si$) gauge fields:
\bea
\vec{E}^a&=&-\pd^0\vec{A}^a-\div\si^a+gf^{abc}\vec{A}^b\si^c
=-\pd^0\vec{A}^a-\vec{D}^{ac}[\vec{A}]\si^c,\nonumber\\
B_i^a&=&\e_{ijk}\left[\nabla_jA_k^a-\frac{1}{2}gf^{abc}A_j^bA_k^c\right]
\eea
and where
\be
\vec{D}_x^{ab}[\vec{A}]=\de^{ab}\div_x-gf^{acb}\vec{A}_x^c
\ee
is the spatial covariant derivative.  $\cd\Phi$ denotes the functional 
integral measure for all field types that may be present.  The action is 
invariant under local (and global), infinitesimal gauge transformations of 
the form:
\bea
\vec{A}_x^{a\th}&=&\vec{A}_x^a+\frac{1}{g}\vec{D}_x^{ab}[\vec{A}]\th_x^b,
\nonumber\\
\si_x^{a\th}&=&\si_x^a-\frac{1}{g}\pd_x^0\th_x^a-f^{acb}\si_x^c\th_x^b.
\label{eq:th}
\eea
Notice that the functional integral measure, $\cd\Phi$, includes all those 
configurations related by such gauge transformations and thus contains the 
integration over the gauge group.  Since the action is gauge invariant, 
integrating over the gauge group produces a (divergent) global factor which, 
in principle, can be absorbed into the normalization of the functional 
integral and in itself is harmless.  However, inconsistent integration over 
the gauge group does become problematic when calculating gauge-variant 
Green's functions from the functional integral: the integration over the 
gauge group averages such quantities to zero.  For example, in the free 
(non-interacting) theory it is necessary to fix the gauge in order to be 
able to invert the quadratic part of the action such that particle 
propagation may be properly defined.  We fix the gauge by applying the 
Faddeev-Popov technique.  This involves inserting the following identity 
($\th_x$ is the parameter associated with the gauge transformation, see 
\eq{eq:th}) into the above functional integral:
\be
\openone=\int\cd\theta_x\de(F[A])\mbox{det}\left[M^{ba}(y,x)\right],
\;\;\;\;M^{ba}(y,x)=\left.\frac{\de F^a[A^{\th}(x)]}{\de\th_y^b}\right|_{F=0}.
\label{eq:fpid}
\ee
We are concerned here with Coulomb gauge:
\be
F^a[A]\equiv\s{\div}{\vec{A}^a}=0
\label{eq:coul}
\ee
and for which the Faddeev-Popov kernel reads
\be
M^{ba}(y,x)\sim-\s{\div_x}{\vec{D}_x^{ab}[\vec{A}]}\de(y-x).
\ee
There are caveats to the identity, \eq{eq:fpid}, above.  When the gauge 
fixing is incomplete, zero-modes of the Faddeev-Popov operator will arise, 
i.e., $\th$ or $\vec{A}$ are such that $F[A^{\th}]=F[A]$ and for 
Coulomb gauge, \eq{eq:coul}, there are two cases:
\begin{enumerate}
\item $-\s{\div_x}{\vec{D}_x^{ab}[\vec{A}]}\th_x^b=0$,
 Gribov copies for $\vec{A}\neq0$ (generated by spatially and temporally 
 dependent gauge transformations)
\item $-\s{\div_x}{\vec{D}_x^{ab}[\vec{A}]}\th^b(t)=0$,
 temporal zero-modes (generated by temporally dependent but spatially 
 independent gauge transformations).
\end{enumerate}
[We implicitly include global ($\th=\mbox{constant}$) transformations into the 
latter category.]  Clearly, in both these cases, $M$ has zero eigenvalues and 
consequently the Faddeev-Popov determinant also vanishes, 
violating \eq{eq:fpid} since the left-hand side cannot vanish.  We must 
therefore modify the formalism to account for this and we write
\be
\openone=
\int\cd\ov{\th}_x\de(F[A])\ov{\mbox{det}}\left[M^{ba}(y,x)\right],
\ee
where $\cd\ov{\th}_x$ explicitly excludes any spatially independent 
$\th(t)$ and
\be
\ov{\mbox{det}}\left[M^{ba}(y,x)\right]=
\mbox{det}\left[M^{ba}(y,x)\right]_{-\s{\div}{\vec{D}}\th\neq0}
\ee
is the determinant with the zero-modes of the operator (temporal or Gribov 
copy) removed.  We can now write our functional integral as
\be
Z=\int\cd\Phi\de(\s{\div}{\vec{A}^a})
\ov{\mbox{det}}\left[-\s{\div}{\vec{D}}\right]
\exp{\left\{\imath\cs_{YM}\right\}}
\ee
where it is understood that the functional integration measure, $\cd\Phi$, 
still contains the integration over the full gauge group.  The direction(s) 
in group space corresponding to the zero-modes of the Faddeev-Popov kernel 
are still explicitly present within the functional integration and may still 
cause problems.

To proceed, we convert to the first order formalism 
\cite{Zwanziger:1998ez,Watson:2006yq}.  
We introduce an auxiliary vector field ($\vec{\pi}$) via the following 
identity:
\be
\exp{\left\{\imath\int\dx{x}\frac{1}{2}\s{\vec{E}^a}{\vec{E}^a}\right\}}=
\int\cd\vec{\pi}
\exp{\left\{\imath\int\dx{x}\left[
-\frac{1}{2}\s{\vec{\pi}^a}{\vec{\pi}^a}-\s{\vec{\pi}^a}{\vec{E}^a}
\right]\right\}}.
\ee
The $\vec{\pi}$-field is further split into components using the identity
\be
\mbox{const}=\int\cd\phi\,\cd\ta\,
\exp{\left\{-\imath\int\dx{x}\ta^a
\left(\s{\div}{\vec{\pi}^a}+\nabla^2\phi^a\right)\right\}},
\ee
changing variables $\vec{\pi}=\vec{\pi}-\div\phi$ and integrating out the 
Lagrange multiplier field, $\ta$, to arrive at the form
\be
Z=\int\cd\Phi\de(\s{\div}{\vec{A}^a})\de(\s{\div}{\vec{\pi}^a})
\ov{\mbox{det}}\left[-\s{\div}{\vec{D}}\right]\exp{\left\{\imath\cs\right\}}
\ee
where
\be
\cs=\int\dx{x}\left[-\frac{1}{2}B^2-\frac{1}{2}\pi^2
-\frac{1}{2}(\nabla\phi)^2+\s{\vec{\pi}^a}{\pd^0\vec{A}^a}
+\si^a\left(\s{\div}{\vec{D}^{ab}}\phi^b+g\hat{\ro}^a\right)\right]
\ee
and with the color charge of the gauge field
\be
\hat{\ro}^a=f^{ade}\s{\vec{A}^d}{\vec{\pi}^e}.
\label{eq:eff}
\ee
The advantage of the first order formalism is, of course, that the action is 
linear in $\si$ and this field can be integrated out to give
\be
Z=\int\cd\Phi\de(\s{\div}{\vec{A}^a})\de(\s{\div}{\vec{\pi}^a})
\ov{\mbox{det}}\left[-\s{\div}{\vec{D}}\right]
\de\left(\s{\div}{\vec{D}^{ab}}\phi^b+g\hat{\ro}^a\right)
\exp{\left\{\imath\cs'\right\}}
\label{eq:funint}
\ee
where
\be
\cs'=\int\dx{x}\left[-\frac{1}{2}B^2-\frac{1}{2}\pi^2
-\frac{1}{2}(\nabla\phi)^2+\s{\vec{\pi}^a}{\pd^0\vec{A}^a}\right].
\ee

The $\de$-functional constraint on the scalar field $\phi$ is the functional 
expression of Gauss' law.  To resolve the constraint and eliminate the 
functional integration over the scalar field $\phi$, we first must 
investigate the Faddeev-Popov operator a little more closely.  The 
eigenvalue equation for the operator is
\be
-\s{\div}{\vec{D}_x^{ab}[\vec{A}]}\vph_n^b(\vec{x};t)=\la_n\vph_n^a(\vec{x};t)
\ee
with a complete orthonormal basis ${\vph_n^a(\vec{x};t)}$ satisfying
\be
\int\dx{x}\vph_m^{*a}(\vec{x};t)\vph_n^a(\vec{x},t)=\de_{mn}.
\ee
However, since the time dependence of the operator is only implicit (within 
the spacetime dependent field $\vec{A}_x$), the time argument is only a label 
(in the sense that there is a different spatial operator at each time).  
We could state that the eigenvalue equation is evaluated at a specific time 
and for which the eigenfunctions are spatially orthonormalized at this 
time.  In this case, the eigenvalues become implicitly time dependent in 
that they refer to the eigenvalue equation at a particular time, $t$:
\be
-\s{\div}{\vec{D}_x^{ab}[\vec{A}]}\vph_{tn}^b(\vec{x})
=\la_{tn}\vph_{tn}^a(\vec{x}),\;\;\;\;
\int d^3x\vph_{tm}^{*a}(\vec{x})\vph_{tn}^a(\vec{x})=\de_{mn}.
\label{eq:orth0}
\ee
We use the convention that $\la_{tn=0}=0$ denotes the collection of zero 
modes of the Faddeev-Popov operator.  The temporal zero-modes are spatially 
constant and for $SU(N_c)$, there are $N_c^2-1$ such linearly independent 
eigenvectors in the color space which when necessary, we label by an 
additional index $\mu$: $\vph_{t0}^b(\vec{x})=\vph_{t0\mu}^b$ (the adjoint 
color index, $b$ here, labels the component of the $\mu$-th vector).  Using 
the complete orthonormal basis, \eq{eq:orth0}, we can expand the field, 
$\phi$, and color charge, $\hat{\ro}$, as series (index $\mu$ is implicit 
within the zero modes):
\be
\phi_x^a=\sum_{n=0}^{\infty}b_n(t)\vph_{tn}^a(\vec{x}),\;\;\;\;
\hat{\ro}_x^a=\sum_{n=0}^{\infty}a_n(t)\vph_{tn}^a(\vec{x})
\ee
where the coefficients are time dependent and, in particular,
\be
a_{0\mu}(t)\sim\int d^3x\vph_{t0\mu}^{*a}(\vec{x})\hat{\ro}_x^a.
\label{eq:rho}
\ee
Having identified the zero-modes, albeit formally, we may define the inverse 
Faddeev-Popov operator in their absence, $\ov{M}$:
\be
\left[-\s{\div}{\vec{D}}\right]\ov{M}\Psi_x=
\ov{M}\left[-\s{\div}{\vec{D}}\right]\Psi_x=\Psi_x
\ee
such that
\be
\mbox{det}\ov{M}=\prod_{n\neq0}\la_{tn}^{-1}=
\ov{\mbox{det}}\left[-\s{\div}{\vec{D}}\right]^{-1}.
\ee

Returning to the functional integral over $\phi$ within the expression, 
\eq{eq:funint}, we may now write
\bea
\lefteqn{
\int\cd\phi\de\left(\s{\div}{\vec{D}^{ab}}\phi^b+g\hat{\ro}^a\right)
\exp{\left\{\imath\int\dx{x}\left[-\frac{1}{2}(\nabla\phi)^2\right]\right\}}
}\nonumber\\
&=&\int\left[
\prod_{n=0}^{\infty}db_n(t)\de\left(-b_n(t)\la_{tn}+ga_n(t)\right)\right]
\exp{\left\{\imath\int\dx{x}\frac{1}{2}
\sum_{m,n=0}^{\infty}b_m^*(t)b_n(t)
\vph_{tm}^{*a}(\vec{x})\nabla_x^2\vph_{tn}^a(\vec{x})\right\}}\nonumber\\
&=&\prod_{\mu}\de\left(ga_{0\mu}(t)\right)
\int_{-\infty}^{\infty}db_{0\mu}(t)\int\left[
\prod_{n=1}^{\infty}db_n(t)\de\left(-b_n(t)\la_{tn}+ga_n(t)\right)\right]
\times\nonumber\\&&
\exp\left\{\frac{\imath}{2}\int\dx{x}\left[\left|b_{0\mu}(t)\right|^2
\vph_{t0\mu}^{*a}(\vec{x})\nabla_x^2\vph_{t0\mu}^a(\vec{x})
+2\sum_{n=1}^{\infty}b_{0\mu}^*(t)b_n(t)
\vph_{t0\mu}^{*a}(\vec{x})\nabla_x^2\vph_{tn}^a(\vec{x})
\right.\right.\nonumber\\&&\left.\left.
+\sum_{m,n=1}^{\infty}b_m^*(t)b_n(t)
\vph_{tm}^{*a}(\vec{x})\nabla_x^2\vph_{tn}^a(\vec{x})\right]\right\}
\nonumber\\
&=&\prod_{\mu}\de\left(ga_{0\mu}(t)\right)
\ov{\mbox{det}}\left[-\s{\div}{\vec{D}}\right]^{-1}
\int_{-\infty}^{\infty}db_{0\mu}(t)\times
\nonumber\\&&
\exp\left\{\frac{\imath}{2}\int\dx{x}\left[\left|b_{0\mu}(t)\right|^2
\vph_{t0\mu}^{*a}(\vec{x})\nabla_x^2\vph_{t0\mu}^a(\vec{x})
+2g\sum_{n=1}^{\infty}b_{0\mu}^*(t)\frac{a_n(t)}{\la_{tn}}
\vph_{t0\mu}^{*a}(\vec{x})\nabla_x^2\vph_{tn}^a(\vec{x})
\right.\right.\nonumber\\&&\left.\left.
+g^2\sum_{m,n=1}^{\infty}\frac{a_m^*(t)a_n(t)}{\la_{tm}^*\la_{tn}}
\vph_{tm}^{*a}(\vec{x})\nabla_x^2\vph_{tn}^a(\vec{x})\right]\right\}.
\eea
If we neglect the (spatially dependent) Gribov copies, the temporal zero 
modes can be completely eliminated by noting that $\div_x\vph_{t0\mu}^a=0$, 
such that the integrals over the $b_{0\mu}(t)$ are overall (divergent) 
constants which can be absorbed into the normalization \footnote{One can 
appreciate the nature of the full Gribov problem very clearly within this 
context.  The eigenfunctions, $\vph_{tn}^a(\vec{x})$, of the Faddeev--Popov 
operator in this case are implicitly dependent on the field configurations 
$\vec{A}(x)$ which, in the complete functional integral, are being 
integrated over.  Without the simplification as in the temporal case 
considered in this study, one is left with a seemingly intractable 
problem.}.  Thus, using \eq{eq:rho}
\bea
\lefteqn{
\int\cd\phi\de\left(\s{\div}{\vec{D}^{ab}}\phi^b+g\hat{\ro}^a\right)
\exp{\left\{\imath\int\dx{x}\left[-\frac{1}{2}(\nabla\phi)^2\right]\right\}}
}\nonumber\\
&=&\prod_{\mu}\de\left(g\vph_{t0\mu}^{*a}\int  d^3x\hat{\ro}_x^a\right)
\ov{\mbox{det}}\left[-\s{\div}{\vec{D}}\right]^{-1}
\exp\left\{\frac{\imath}{2}\int\dx{x}\left[g^2\sum_{m,n=1}^{\infty}
\frac{a_m^*(t)a_n(t)}{\la_{tm}^*\la_{tn}}
\vph_{tm}^{*a}(\vec{x})\nabla_x^2\vph_{tn}^a(\vec{x})\right]\right\}.
\eea
Now, because the temporal zero-mode is also a zero-mode of the Laplacian, we 
can infer that the ratio $\div\vph_{t0\mu}/\la_{t0}$ is finite and that if we 
multiply by $a_{0\mu}(t)=0$ then the product vanishes.  Thus, the sum in the 
exponential can be extended to include the zero-modes without reintroducing 
any ambiguities (we are, after all, only adding zero) and we have the form
\bea
\lefteqn{
\int\cd\phi\de\left(\s{\div}{\vec{D}^{ab}}\phi^b+g\hat{\ro}^a\right)
\exp{\left\{\imath\int\dx{x}\left[-\frac{1}{2}(\nabla\phi)^2\right]\right\}}
}\nonumber\\
&\!\!\!\!=&\!\!\!\!
\prod_{\mu}\de\left(g\vph_{t0\mu}^{*a}\int d^3x\hat{\ro}_x^a\right)
\ov{\mbox{det}}\left[-\s{\div}{\vec{D}}\right]^{-1}
\exp\left\{\frac{\imath}{2}\int\dx{x}\left[g^2\hat{\ro_x^a}
\left[-\s{\div_x}{D_x^{ab}[\vec{A}]}\right]^{-1}\nabla^2
\left[-\s{\div_x}{D_x^{bc}[\vec{A}]}\right]^{-1}\hat{\ro_x^c}\right]\right\}
\label{eq:phiint}
\eea
where the quantities in the exponent formally include the zero-modes.  As 
demonstrated, these zero-modes play no role other than to allow us to write 
the action conventionally in terms of the full fields; however, nontrivially, 
these full fields have been shown to be well-defined insofar as the temporal 
zero-modes are concerned.  Since the $\vph_{t0\mu}^a$ are ($N_c^2-1$) linearly 
independent vectors in the adjoint color space, we also have that
\be
\prod_{\mu}\de\left(g\vph_{t0\mu}^{*a}\int d^3x\hat{\ro}_x^a\right)
\rightarrow\prod_{a}\de\left(\int d^3x\hat{\ro}_x^a\right).
\ee

Returning to the original functional integral, \eq{eq:funint}, and 
substituting in the above result for the $\phi$ integration, \eq{eq:phiint}, 
we see immediately that the modified Faddeev--Popov determinants cancel, 
leaving
\be
Z=\int\cd\Phi\de(\s{\div}{\vec{A}^a})\de(\s{\div}{\vec{\pi}^a})
\de\left(\int d^3x\hat{\ro}_x^a\right)\exp{\left\{\imath\cs''\right\}}
\label{eq:fnunt}
\ee
with
\be
\cs''=\int\dx{x}\left[
-\frac{1}{2}B^2-\frac{1}{2}\pi^2+\s{\vec{\pi}^a}{\pd^0\vec{A}^a}
+\frac{1}{2}g^2\hat{\ro_x^a}\left[-\s{\div_x}{D_x^{ab}[\vec{A}]}\right]^{-1}
\nabla^2\left[-\s{\div_x}{D_x^{bc}[\vec{A}]}\right]^{-1}\hat{\ro_x^c}\right].
\label{eq:effact}
\ee
Given the definition of the color charge $\hat{\ro}^a$, 
\eq{eq:eff}, it is clear that the argument of its $\de$-functional 
constraint ($\int d^3xf^{abc}\s{\vec{A}^b}{\vec{\pi}^c}$) must vanish at 
each time, $t$.  Further, the vanishing of the charge is invariant under 
temporal gauge transformations since (using the fundamental representation 
for the colored fields and denoting the spatially independent gauge 
transformation in this representation $U_t$)
\be
0=\int d^3x\ro=\int  d^3x\left[\vec{A},\vec{\pi}\right]
\rightarrow\int d^3x\left[U_t\vec{A}U_t^\dag,U_t\vec{\pi}U_t^\dag\right]
=U_t\int d^3x\left[\vec{A},\vec{\pi}\right]U_t^\dag=0.
\ee
The action, $\cs''$ given by \eq{eq:effact}, is however no longer invariant 
under such temporal transformations (courtesy of the 
$\s{\vec{\pi}^a}{\pd^0\vec{A}^a}$ term which remains after we have 
integrated out the $\si$-field) and the temporal zero-modes of our new 
functional integral, \eq{eq:fnunt}, do not exist.  This functional integral 
is thus fully gauge-fixed (it contains no more zero-modes, except those Gribov 
copies that we ignore here) and constrains the conserved, total color charge 
to be vanishing:
\be
\int d^3x\hat{\ro}_x^a=\int d^3xf^{abc}\s{\vec{A}^b}{\vec{\pi}^c}=0.
\ee
The appearance of the $\de$-functional constraint in \eq{eq:fnunt}, which 
ensures the vanishing of the total color charge is the main result of this 
work.  Given that the action, \eq{eq:effact}, is unchanged from the form 
originally derived in \cite{Zwanziger:1998ez}, the consideration of the 
zero-modes leads to the further observation that the dynamics of the theory 
are unaltered.

If quarks were to be included in the original action, then the color charge 
acquires an extra component:
\be
\hat{\ro}_q^a=\ov{q}T^a\ga^0q
\ee
where $T^a$ is the (Hermitian) generator of the gauge group, $q$ is the 
quark field and $\ov{q}$ its conjugate.  The total charge is constrained as 
before, i.e.,
\be
\int d^3x\left[\hat{\ro}^a+\hat{\ro}_q^a\right]=0.
\ee
The inclusion of quarks into the first order formalism is currently being 
studied \cite{carina}.

\section{Summary and conclusions}
\setcounter{equation}{0}
The temporal (and global) zero-modes of the Faddeev--Popov operator for 
Coulomb gauge, continuum Yang--Mills theory within the first order formalism 
have been studied.  The explicit separation of such modes leads to a 
well-defined total color charge that is conserved and vanishing at all times.

The appearance of the total color charge is interesting for 
several reasons.  Firstly, that the total charge of the system is conserved 
and vanishing is a necessary condition for confinement --- one could not 
expect only color singlet hadrons to emerge if the system were not so.  As 
such, the properties of this charge may be regarded as a proof that at the 
least, the total system (i.e., the universe) is colorless.  This does not 
however say much about the confinement of quarks and gluons, or the observed 
spectrum of hadrons.  That being said though, whilst the temporal zero-modes 
lead only to a total conserved color charge, one may speculate about the 
role of the Gribov copies (i.e., the spatially dependent zero-modes) in 
confinement.  This is the premise of the Gribov--Zwanziger confinement 
scenario \cite{Gribov:1977wm,Zwanziger:1995cv,Zwanziger:1998ez} and the 
above analysis intuitively supports this.

Second, given that the temporal zero-modes drop out of the functional 
integral without modification of the effective, gauge-fixed action despite 
the fact that the (Coulomb) gauge-fixing is incomplete leads to the 
conclusion that no further gauge-fixing considerations are necessary and that 
Coulomb gauge is a well-defined choice of gauge insofar as the dynamics are 
concerned (leaving aside the definition of the physical state space).  
Further, since such temporal zero-modes are associated with the ghost energy 
divergence problem it is clear that these divergences must cancel in final 
(physically meaningful) expressions, at least in principle (although they 
may occur in individual components).

Thirdly, Gauss' law (which here appears as the $\de$-functional constraint 
on the auxiliary $\phi$-field in \eq{eq:funint} and arises when one 
integrates out the temporal, $\si$-field) plays a pivotal role in the 
extraction of the zero-modes and therefore in both the definition 
(and conservation) of the total color charge and in resolving the 
incompleteness of the gauge (see the previous paragraph).  This is not 
surprising, since Gauss' law defines the charge via the generator of gauge 
transformations \cite{Reinhardt:2008ek}.  In the Hamiltonian approach to 
Yang--Mills theory \cite{Christ:1980ku}, the imposition of Gauss' law as a 
constraint on the physical state space ensures gauge invariance in exactly 
this way.  The resolution of Gauss' law in Coulomb gauge results in the 
explicit appearance of the nonabelian Coulomb potential in the Hamiltonian 
(and also here as the last term in the effective action, \eq{eq:effact}).

Fourthly, whilst the first order formalism has been employed here, the 
manipulations required to construct this formalism from the more usual 
second order formalism comprise identities that merely serve to linearize 
the action with respect to the temporal gauge field, $\si$.  Thus, what is 
true in the first order formalism will also be true in the second order 
formalism, although it will be manifested differently.  The eventual form of 
the functional integral, \eq{eq:fnunt}, with the effective action, 
\eq{eq:effact}, here are non-local expressions and as such are not amenable 
to renormalization (that the first order formalism is not 
\emph{multiplicatively} renormalizable is known \cite{Zwanziger:1998ez}).  
In the standard, local version of the second order formalism (i.e., where the 
Faddeev-Popov determinant is expressed in terms of Grassmannian ghost fields 
etc.) which appears to be locally renormalizable (although there is as yet no 
conclusive proof) and therefore more suited to calculation, it may indeed not 
be possible to write the constraint that the total color charge is conserved 
and vanishing in convenient form but this does not mean that this is not 
true --- quite the opposite.  In any event, the dynamics of the theory are 
unaltered and in particular from a practical standpoint, the incomplete 
gauge-fixing poses no fundamental problem and the ghost energy divergences 
will cancel in the second order formalism just as in the first.

It is worth pointing out that in principle, the $\vec{\pi}$-field in the 
final expressions, \eq{eq:fnunt} and \eq{eq:effact}, can be integrated out 
(the exponent is at most quadratic in $\vec{\pi}$) and the $\de$-functional 
constraint on the total charge explicitly resolved.  Whether or not the 
resulting expressions have any practical use however, remains to be seen, 
because of their highly non-local nature.

Finally, the gauge invariance of the theory as applied to the Green's 
functions leads to Ward--Takahashi and Slavnov--Taylor identities.  If one 
naively applies a temporal gauge transformation and tries to construct these 
identities, the resulting expressions are meaningless since the functional 
integration measure is not properly defined.  As such, this is the only 
visible effect of leaving the temporal zero-modes of the Faddeev--Popov 
operator in the functional integral.  However, once recognized this poses no 
problem in practice since to derive such identities one generally considers a 
fully spacetime dependent gauge transformation.  The derivation of such 
identities is the focus of present work \cite{current}.

\begin{acknowledgments}
This work has been supported by the Deutsche Forschungsgemeinschaft (DFG) 
under contracts no. DFG-Re856/6-1 and DFG-Re856/6-2.
\end{acknowledgments}

\end{document}